\documentclass[preprint,12pt,authoryear]{elsarticle}

\usepackage{xcolor} 
\usepackage{rotating}
\usepackage{makecell}
\usepackage{xstring}
\usepackage{pbox}
\usepackage{hyperref}
\usepackage{setspace}

\usepackage{amssymb}

\usepackage{booktabs}

\usepackage{multirow}

\newcommand{\beginsupplement}{%
        \setcounter{table}{0}
        \renewcommand{\thetable}{S\arabic{table}}%
        \setcounter{figure}{0}
        \renewcommand{\thefigure}{S\arabic{figure}}%
}

\journal{Medical Image Analysis}

\begin{document}
\doublespacing
\begin{frontmatter}

\title{Automatic quality control of brain T1-weighted magnetic resonance images for a clinical data warehouse}

\author[Inria,SU,ICM,INSERM,CNRS,APHP]{Simona Bottani}
\author[SU,ICM,INSERM,CNRS,APHP,Inria]{Ninon Burgos}
\author[WIND]{Aurélien Maire}
\author[SU,ICM,INSERM,CNRS,APHP,Inria]{Adam Wild}
\author[Neuroradio]{Sebastian Ströer}
\author[SU,ICM,INSERM,CNRS,APHP,Inria,Neuroradio]{Didier Dormont}
\author[SU,ICM,INSERM,CNRS,APHP,Inria]{Olivier Colliot}
\author[]{APPRIMAGE Study Group}
\affiliation[Inria]{organization={Inria, Aramis project-team},
            city={Paris},
            postcode={75013},
            country={France}}

\affiliation[SU]{organization={Sorbonne Université},
            city={Paris},
            postcode={75013},
            country={France}}
\affiliation[ICM]{organization={Institut du Cerveau ‐ Paris Brain Institute ‐ ICM},
            city={Paris},
            postcode={75013},
            country={France}}
\affiliation[INSERM]{organization={Inserm},
            city={Paris},
            postcode={75013},
            country={France}}
\affiliation[CNRS]{organization={CNRS},
            city={Paris},
            postcode={75013},
            country={France}}
\affiliation[APHP]{organization={AP-HP, Hôpital de la Pitié Salpêtrière},
            city={Paris},
            postcode={75013},
            country={France}}
\affiliation[WIND]{organization={AP-HP, WIND department},
            city={Paris},
            postcode={75012},
            country={France}}
\affiliation[Neuroradio]{organization={AP-HP, Hôpital de la Pitié Salpêtrière, Department of Neuroradiology},
            city={Paris},
            postcode={75013},
            country={France}}

\begin{abstract}
Many studies on machine learning (ML) for computer-aided diagnosis have so far been mostly restricted to high-quality research data. Clinical data warehouses, gathering routine examinations from hospitals, offer great promises for training and validation of ML models in a realistic setting. However, the use of such clinical data warehouses requires quality control (QC) tools. Visual QC by experts is time-consuming and does not scale to large datasets. In this paper, we propose a convolutional neural network (CNN) for the automatic QC of 3D T1-weighted brain MRI for a large heterogeneous clinical data warehouse. To that purpose, we used the data warehouse of the hospitals of the Greater Paris area (Assistance Publique-Hôpitaux de Paris [AP-HP]). Specifically, the objectives were: 1) to identify images which are not proper T1-weighted brain MRIs; 2) to identify acquisitions for which gadolinium was injected; 3) to rate the overall image quality. We used 5000 images for training and validation and a separate set of 500 images for testing. In order to train/validate the CNN, the data were annotated by two trained raters according to a visual QC protocol that we specifically designed for application in the setting of a data warehouse. For objectives 1 and 2, our approach achieved excellent accuracy (balanced accuracy and F1-score \textgreater 90\%), similar to the human raters. For objective 3, the performance was good but substantially lower than that of human raters. Nevertheless, the automatic approach accurately identified (balanced accuracy and F1-score \textgreater 80\%) low quality images, which would typically need to be excluded. Overall, our approach shall be useful for exploiting hospital data warehouses in medical image computing. 
\end{abstract}

\begin{graphicalabstract}
    \includegraphics{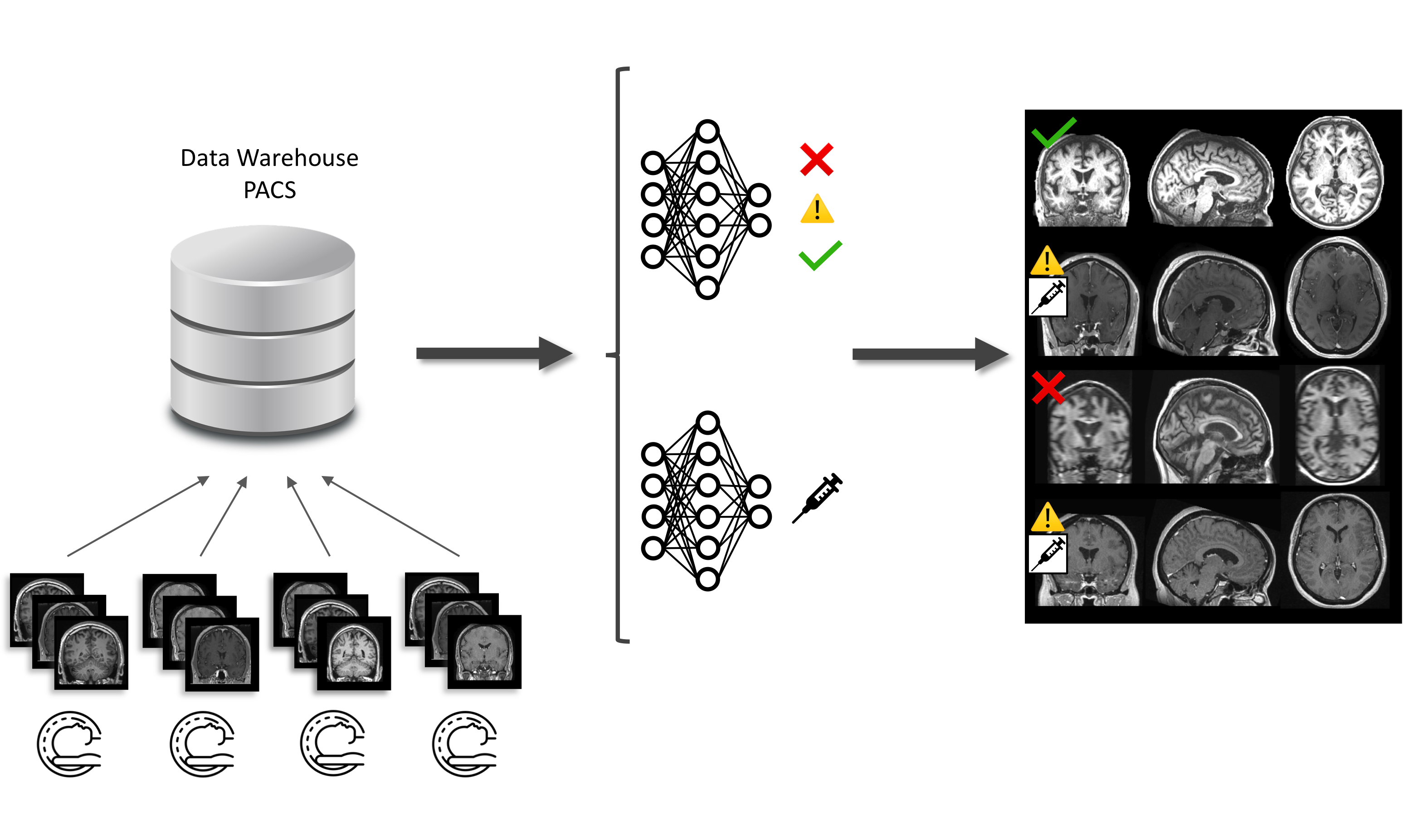}
\end{graphicalabstract}

\begin{highlights}
    \item We propose a framework for the automatic QC of 3D T1w brain MRI for a clinical data warehouse.
    \item We manually labeled 5500 images to train/test different convolutional neural networks.
    \item The automatic approach can identify images which are not proper T1w brain MRIs (e.g. truncated images).
    \item It is able to identify acquisitions for which gadolinium was injected.
    \item It can also accurately identify low quality images.
\end{highlights}


\end{frontmatter}

\section{Introduction}
\label{sec:introduction}

Structural T1-weighted (T1w) magnetic resonance imaging (MRI) is useful for diagnosis of various brain disorders, in particular neurodegenerative diseases \citep{frisoni_clinical_2010, harper2016mri}. They have thus often been used as inputs of machine learning (ML) algorithms for computer-aided diagnosis (CAD) \citep{falahati2014multivariate, koikkalainen2016differential, rathore2017review, burgos2020machine}. 

\begin{figure}[!t]
\begin{center}
    \includegraphics[width=1\linewidth]{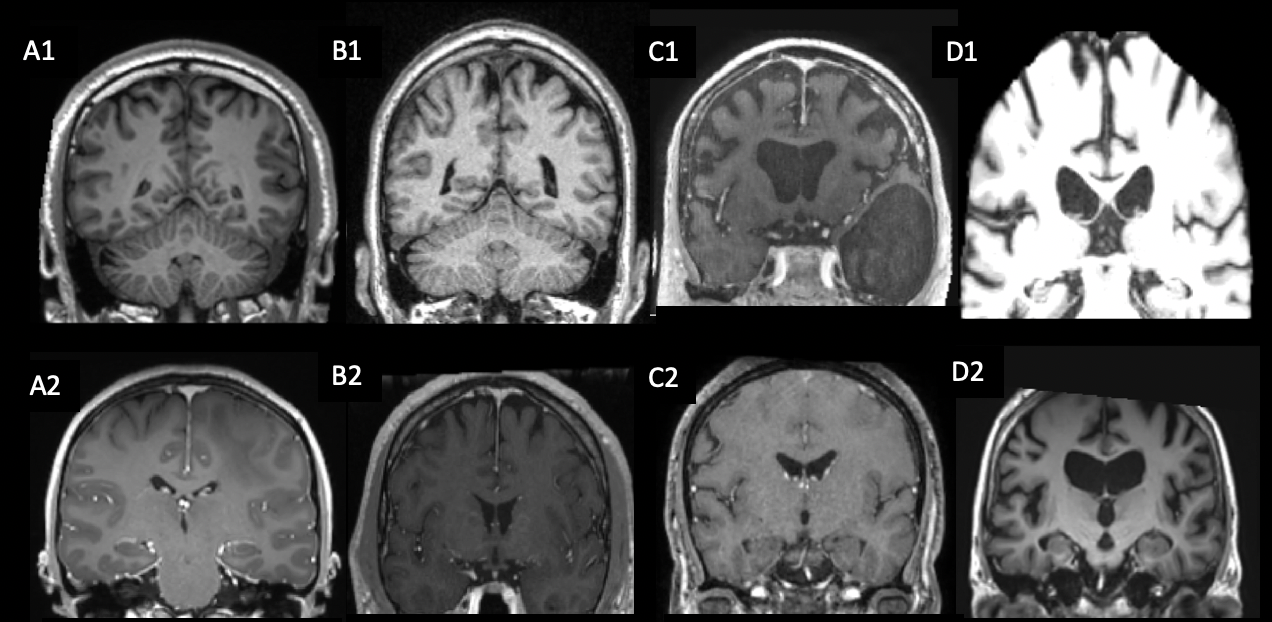}
    \caption{Examples of T1w brain images from the clinical data warehouse and the corresponding labels. A1: Image of good quality (tier 1), without gadolinium; A2: Good quality (tier 1), with gadolinium; B1: Medium quality (tier 2), without gadolinium (noise grade 1); B2: Medium quality (tier 2), with gadolinium (contrast grade 1); C1: Bad quality (tier 3), without gadolinium (contrast grade 2, motion grade 2); C2: Bad quality (tier 3), with gadolinium (contrast grade 2, motion grade 1); D1: Straight rejection (segmented); D2: Straight rejection (cropped).}
    \label{fig:labelbrain}
\end{center}
\end{figure}

Most ML methods are trained and validated on high-quality research data \citep{noor2019detecting,choi2019deep,punjabi2019neuroimaging}: protocols for image acquisition are standardized and a strict quality control is applied \citep{jack2008alzheimer,littlejohns2020uk}. However, to be applied in the clinic, ML methods need to be validated on clinical routine images. In recent years, hospitals have constituted clinical data warehouses that can contain medical images from 100,000-1,000,000 patients \citep{daniel2020hospital,amara2020design}. The quality of such images can greatly vary (see Figure~\ref{fig:labelbrain}), since the acquisition protocols are not standardized, scanners may not be recent and patients may have moved during the acquisition. All these factors can prevent algorithms from working properly \citep{reuter2015head,gilmore2019variations}. Quality control (QC) is thus a fundamental step before training and evaluating ML approaches on clinical routine data. 

Manual QC takes time and is thus not always doable, especially in the context of ML-based CAD, where a large number of training samples is needed. Typically, clinical data warehouses can contain hundreds of thousands of samples. Even if web-based systems facilitate annotation \citep{kim2019loni,keshavan2018mindcontrol}, the task remains unfeasible for very large datasets. In this context, automatic QC is needed. 

Several works have been proposed to enable automatic QC. The Preprocessed Connectomes Project developed a Quality Assessment Protocol\footnote{\url{http://preprocessed-connectomes-project.org/quality-assessment-protocol}}. The package enables the extraction of several image quality metrics (IQMs) such as the signal-to-noise ratio, the contrast-to-noise ratio or the volume of the gray and white matter. IQMs are then compared to a normative distribution obtained from three research datasets, ABIDE \citep{di2014autism}, CoRR\footnote{\url{http://fcon_1000.projects.nitrc.org/indi/CoRR/html/index.html}} and NFB\footnote{\url{http://fcon_1000.projects.nitrc.org/indi/enhanced/}}. 
In the same spirit, we find \citep{esteban2017mriqc,alfaro2018image,raamana2020visual}. These approaches propose to use the IQMs as input of a classifier for automatic QC.
\cite{esteban2017mriqc} and \cite{alfaro2018image} developed a pipeline for the automatic QC of 3D brain T1w MRI, the first has the advantage to be an open source software (called MRIQC). \cite{raamana2020visual} developed another open source software called VisualQC whose aim is the visualisation and the rating of the Freesurfer cortical segmentation output.
The pipelines proposed by these works are very extensive as they require registration and segmentation steps to extract features. It is not possible to assume a priori that these steps will perform well with a new unseen clinical dataset. On the contrary, it is likely that the segmentation will fail for the lowest quality images, thus making it impossible to apply the QC tool. Moreover, the extracted features may not be representative of the problems affecting clinical routine data. As proposed by \cite{sujit2019automated}, convolutional neural networks (CNNs) are a good option for automatic QC because they can learn features without knowing a priori which are the most adapted. A further limitation of these works is that they rely on images acquired following a well-defined research protocol. The pipeline presented in \citep{alfaro2018image} was developed for the large, but well-standardized, UK Biobank dataset containing mostly healthy volunteers. \cite{esteban2017mriqc} and \cite{sujit2019automated} trained their algorithms on ABIDE, a research multicenter study including patients with autism and control subjects and used another research dataset for testing. Thus, to the best of our knowledge, there is currently no automatic QC approach dedicated to large clinical datasets.

Our work was done using a clinical data warehouse. It assembles all MRI data from all hospitals of the greater Paris area. Images come from different sites and different machines with no homogenization on the parameters. Their acquisition cover several decades. The patient may have any disease for which a brain MRI exam is required. All these factors are not present in the approaches already proposed in the literature: even when images come from different sites, the acquisition protocol is harmonized, the number of machines is limited and they are usually acquired within a few years, avoiding intrinsic problems of quality due to the progress in the technology. Additionally, the presence of different diseases such as neurodegenerative diseases, stroke, multiple sclerosis, or brain tumours, is typical of clinical datasets: they can strongly alter the structure of the brain and it may be difficult to use a specific set of features to characterize the quality of the images independently of the disease. In addition, due to security reasons, images from the data warehouse cannot be uploaded to a web server and we had to work in a restricted IT environment \citep{daniel2020hospital}.

The objective of our work was to develop a method for the automatic QC of T1w brain MRI in large clinical data warehouses. The specific objectives were to: 1) discard images which are not proper T1w brain MRI; 2) identify images with gadolinium; 3) recognise images of bad, medium and good quality. We used 5000 images for training/validation and 500 for testing. To train/validate the models, the data were annotated by two trained raters. To that purpose, we introduced an original visual QC protocol that is applicable to clinical data warehouses.

\section{Materials and methods}

\subsection{Dataset description}
This work relies on a large clinical routine dataset containing all the T1w brain MR images of adult patients scanned in hospitals of the Greater Paris area (Assistance Publique-Hôpitaux de Paris [AP-HP]). The data were made available by the data warehouse of the AP-HP and the study was approved by the Ethical and Scientific Board of the AP-HP.
According to French regulation, consent was waived as these images were acquired as part of the routine clinical care of the patients.

The images were selected according to DICOM attributes. A first query on the PACS was performed to list the DICOM attributes corresponding to MRI. For all the MR images, we listed the ``series descriptions", ``body parts examined", and ``study descriptions" DICOM attributes. A neuroradiologist manually selected all the attribute values that may refer to 3D T1w brain MRI (e.g. ``T1 EG 3D MPR", ``SAG 3D BRAVO", ``3D T1 EG MPRAGE", ``IRM cranio", ``Brain T1W/FFEGADO"). He selected 3736 relevant attribute values. Relevant attribute values were manually selected as several DICOM tags are manually filled by radiographers, and so may not be homogeneous for the images acquired across the 39 hospitals of the AP-HP during several decades. These attributes were used to select the images of interest.
Among all the 3D T1w brain MRI of the AP-HP, a first batch of about 11,000 images was delivered by the data warehouse. We excluded all the images having less than 40 slices because they correspond to 2D brain images even if the corresponding DICOM attribute refer to 3D. For the present study, we randomly selected 5500 images, corresponding to 4177 patients. The images were acquired on various scanners from four manufacturers: Siemens Healthineers ($n=3752$), GE Healthcare ($n=1710$), Philips ($n=33$) and Toshiba ($n=5$).  Among all the images, 3229 images were acquired with 3 Tesla machines and 2271 with 1.5 Tesla. Table~\ref{tab:machines_name} in Supplementary Material reports all the models present in our dataset with the corresponding magnetic field.

\subsection{Image preprocessing}
The T1w MR images were converted from DICOM to NIfTI using the software dicom2niix \citep{li2016first} and organized using the Brain Imaging Data Structure (BIDS) standard \citep{gorgolewski2016brain}. Images with a voxel dimension smaller than 0.9~mm were resampled using a 3rd-order spline interpolation to obtain 1~mm isotropic voxels.
To facilitate annotations, we applied the following pre-processing using the ‘t1-linear’ pipeline of Clinica \citep{routier_clinica_2021}, which is a wrapper of the ANTs software \citep{avants2014insight}. Bias field correction was applied using the N4ITK method \citep{tustison2010n4itk}. An affine registration to MNI space was performed using the SyN algorithm \citep{avants2008symmetric}. The registered images were further rescaled based on the min and max intensity values, and cropped to remove background resulting in images of size 169$\times$208$\times$179, with 1~mm isotropic voxels \citep{wen2020convolutional}. One should note that we only aimed to obtain a rough alignment and intensity rescaling to facilitate annotation.

\subsection{Manual labeling of the dataset}
In this section, we introduce the visual QC protocol. We describe the different characteristics noted on the images and how we created the final label for the automatic QC. Images were labeled by two trained raters and the annotation protocol was designed with the help of a radiologist.

\subsubsection{Quality criteria}
Five characteristics were manually annotated. The first two (straight rejection and gadolinium) are binary flags, while the other three (motion, contrast and noise) are assessed with a three-level grade.
\begin{itemize}
\item \textbf{Straight rejection (SR)}: images not containing a T1w MRI of the whole brain (for instance images of segmented tissues or truncated images). Note that these images still have DICOM attributes corresponding to T1w brain MRI and thus were not removed through the selection step based on DICOM attributes.
\item \textbf{Gadolinium}: presence of gadolinium-based contrast agent.
\item \textbf{Motion} 0: no motion,  1: some motion but the structures of the brain are still distinguishable, 2: severe motion, the cortical and subcortical structures are difficult to distinguish.
\item \textbf{Contrast} 0: good contrast, 1: medium contrast (gray matter and white matter are difficult to distinguish in some parts of the image), 2: bad contrast (gray matter and white matter are difficult to distinguish everywhere in the brain).
\item \textbf{Noise} 0: no noise, 1: presence of noise that does not prevent identifying structures, 2: severe noise that does prevent identifying structures.
\end{itemize}
Gadolinium injection, motion, contrast and noise were noted for all the images which were not defined as SR.
According to the grades given to the motion, contrast and noise characteristics, we determined three tiers corresponding to images of good, medium and bad quality. The tiers, along with the rules used to defined them, are described in Table~\ref{tab:summary_tier}.

\begin{table}[!h]
\centering
\renewcommand{\arraystretch}{1.5}

    \begin{tabular}{m{15mm} m{40mm} m{60mm}}
        \toprule
        \bfseries Tier & \bfseries Description &\bfseries Determination \textbf{rule}\\
        \hline\hline
        Tier 1 & 3D T1w brain MRI of good quality & Grade 0 for motion, contrast and noise \\
        \hline
        Tier 2 & 3D T1w brain MRI of medium quality & 
        At least one characteristic among motion, contrast and noise with grade 1 and none with grade 2 \\
        \hline
        Tier 3 & 3D T1w brain MRI of bad quality & At least one characteristic among motion, contrast and noise with grade 2 \\
        \bottomrule
    \end{tabular}
    \caption{Description and determination rules of the proposed quality control tiers.}
    \label{tab:summary_tier}
\end{table}

\subsubsection{Annotation set-up}
Our aim was to annotate the largest possible number of images in an efficient manner while being restricted to the environment of the data warehouse which only included a Jupyter notebook and a command-line interface. We thus implemented a graphical interface in a Jupyter notebook. This interface displayed only the central axial, sagittal and coronal slices of the brain. Indeed, loading the whole 3D volume for inspecting all the slices in the data warehouse environment was unfeasible due to the above mentioned restrictions. Specifically, from the NIfTI format, we saved a screenshot of the central slice of each view (sagittal, coronal, axial) in PNG format. This allowed a fast loading of the image to annotate. Each image was labeled by two trained raters. The interface was flexible: it was possible to go back and label again an image, and after the labelling all the characteristics noted were displayed. The procedure was optimized to reduce the workload of the raters to a minimum.

\subsubsection{Consensus label}
The final label used to train and validate the automatic QC is a consensus between the two raters. If the users labeled different image characteristics, we determined a procedure to define a consensus label. We distinguished two types of disagreement: one regarding the SR status and the other one regarding the other characteristics based on which the tiers are assigned. 
When the two raters disagreed on the SR status, we manually set the consensus label: the two raters reviewed the images and decided together to keep the SR label or assign the alternative label. 
In case of disagreement regarding the other characteristics, the consensus was chosen as follows. The objective was to be as conservative as possible: we wanted to retain all the imperfections that may have been seen by one annotator and not by the other. For a given characteristic, the consensus grade was chosen as the maximum of the two grades of the observers. The tier was recomputed accordingly. 

\subsection{Automatic quality control method}
We developed an automatic QC method based on CNNs trained to perform several classification tasks: 1) discard images which were not proper T1w brain MRI (SR: yes vs no)); 2) identify images with gadolinium (gadolinium: yes vs no); 3) differentiate images of bad quality from images of medium and good quality (tier 3 vs tiers 2-1); 4) differentiate images of medium quality from images of good quality (tier 2 vs tier 1).

\subsubsection{Network architecture}
\begin{figure}[!t]
\begin{center}
\includegraphics[width=1\linewidth]{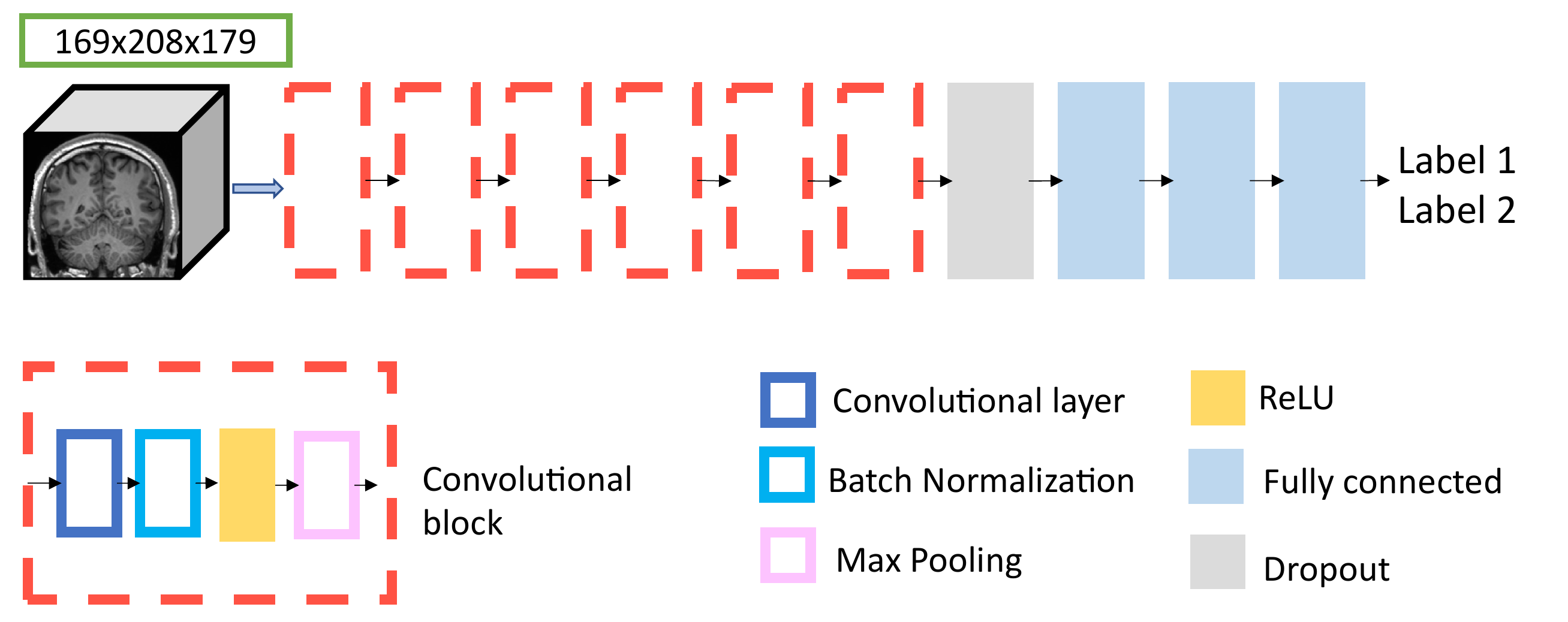}
    \caption{Architecture of the 3D CNN called Conv5\textunderscore FC3. Five convolutional blocks (composed sequentially of a convolutional layer, a batch normalization layer, a ReLU and a max pooling layer) are followed by a dropout and three fully connected layers.}
    \label{fig:conv5fc3}
\end{center}
\end{figure}

The network proposed was composed of five convolutional blocks and of three fully connected layers. The convolutional blocks were made of one convolutional layer, one batch normalization layer, one ReLU and one max pooling. Details about architecture are represented on Figure \ref{fig:conv5fc3}. All the details about the parameters of the layers, i.e. the filter size, the number of filters/neurons,  the stride and the padding size and the dropout rate are in the Supplementary Materials in table \ref{tab:parameters}. In the following, we refer to this architecture as Conv5\textunderscore FC3. The models were trained using the cross entropy loss, which was weighted according to the proportion of images per class for each task. We used the Adam optimizer with a learning rate of 1e-4. We implemented early stopping and all the models were evaluated with a maximum of 50 epochs. The batch size was set to 2. The model with the lowest loss was saved as final model. Implementation was done using Pytorch.
 This architecture has previously been used and validated in \citep{wen2020convolutional}. It is available through the ClinicaDL software available on GitHub: \url{https://github.com/aramis-lab/AD-DL}.

\begin{figure}[!t]
    \begin{center}
        \includegraphics[width=1\linewidth]{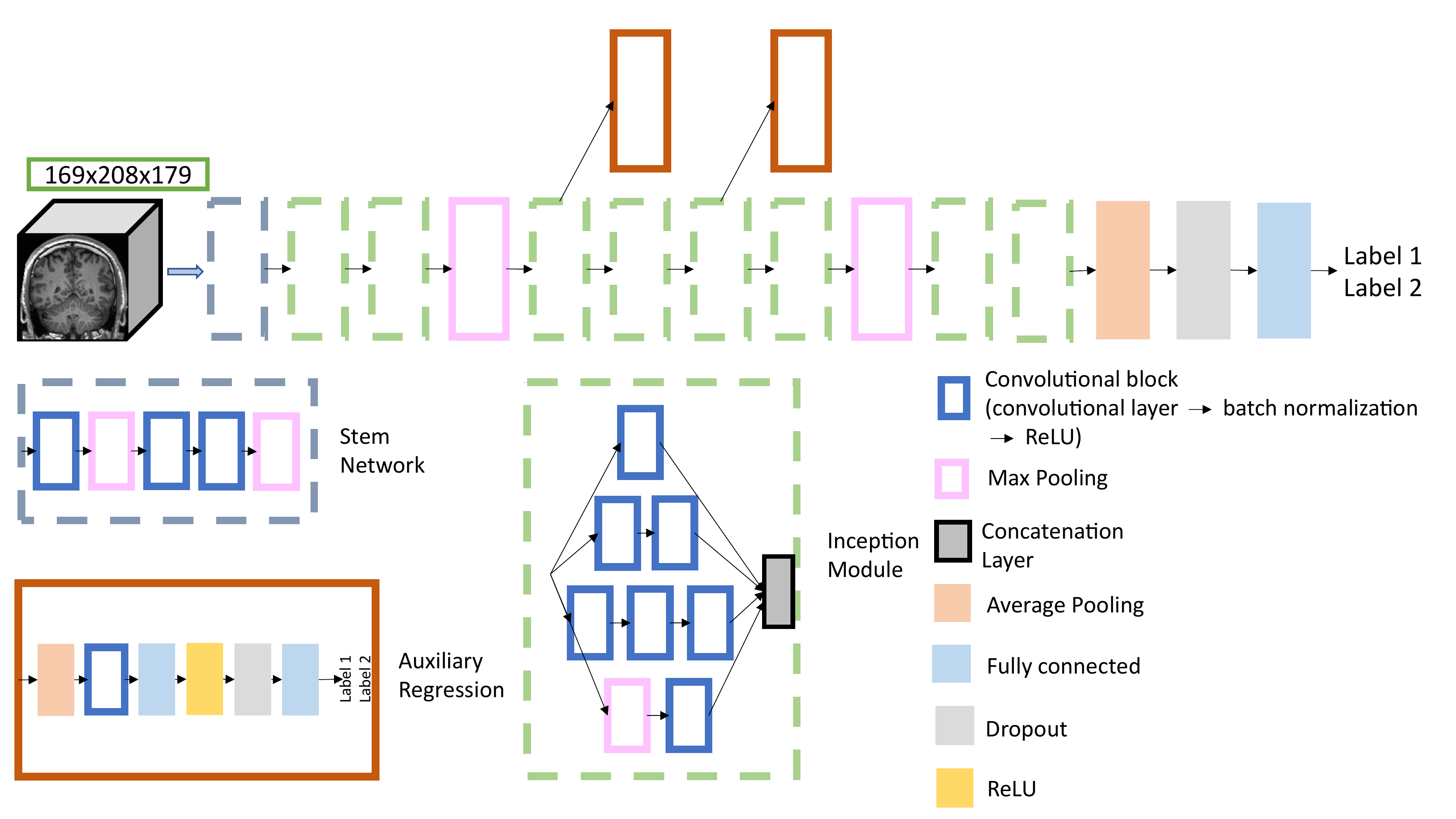}
        \caption{Architecture of the Inception 3D CNN. More information regarding the hyperparameters can be found in \citep{couvy2020ensemble}.}
        \label{fig:inception}
    \end{center}
\end{figure}

\begin{figure}[!t]
    \begin{center}
        \includegraphics[width=1\linewidth]{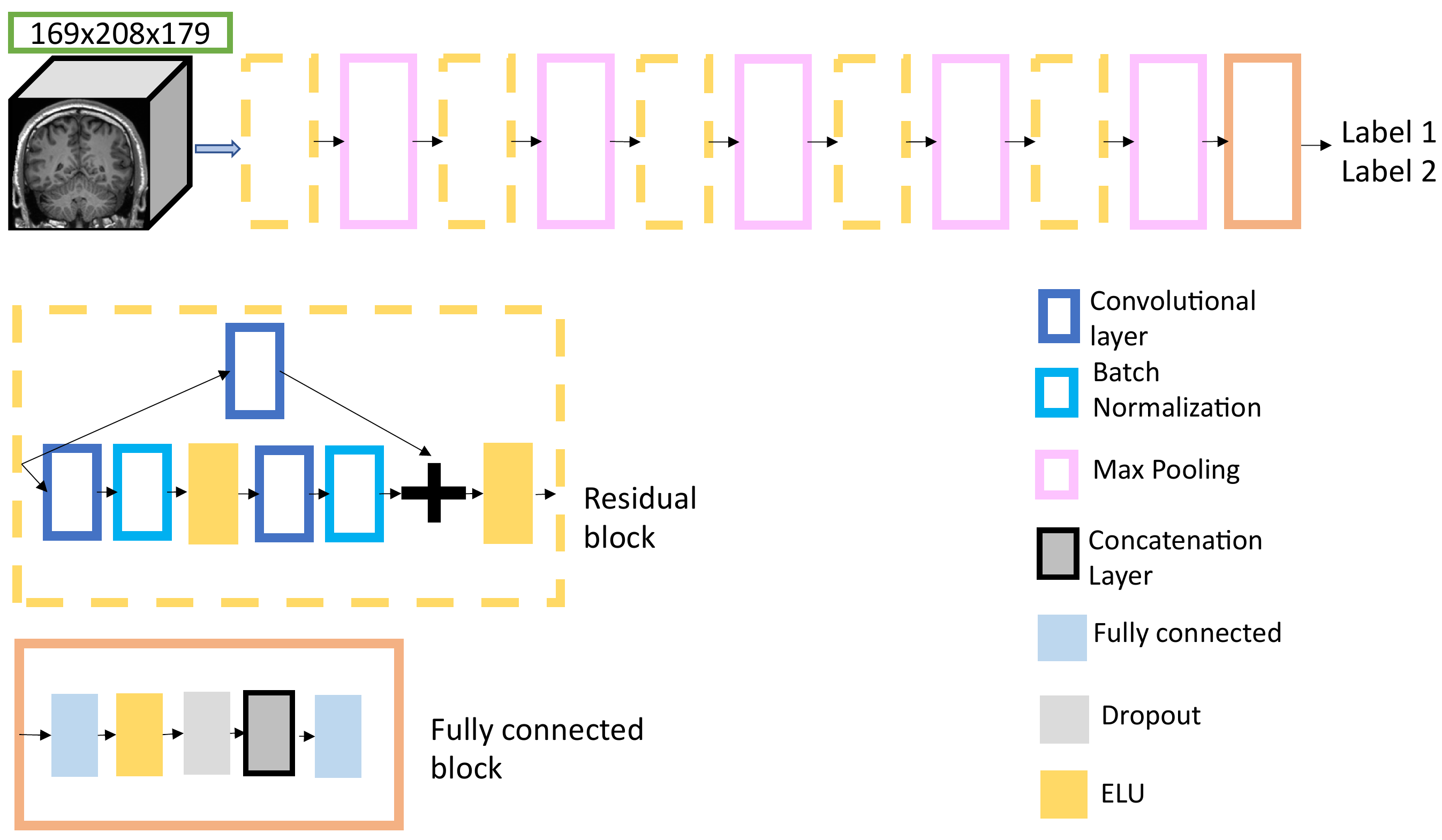}
        \caption{Architecture of the ResNet 3D CNN. More information regarding the hyperparameters can be found in \citep{couvy2020ensemble}.}
        \label{fig:resnet}
    \end{center}
\end{figure}

We compared this network to more sophisticated CNN architectures. In particular, we implemented a modified 3D version of Google's incarnation of the Inception architecture \citep{szegedy2016rethinking}. In addition we also implemented a 3D ResNet (CNN with residual blocks) inspired from \citep{jonsson2019brain}. More details about the architectures are given Figures~\ref{fig:inception} and~\ref{fig:resnet}. Both the Inception and the ResNet models were trained using the cross entropy loss weighted according to the proportion of images per class, the Adam optimizer with a learning rate of 1e-4 and the batch size was set to 2. These two models have been used in \citep{couvy2020ensemble} to predict brain age from 3D T1w MRI. For that specific task, they achieved a higher performance than the 5-layer CNN mentioned above. Their implementation is openly available on GitHub \url{https://github.com/aramis-lab/pac2019} and all the parameters of the CNNs are listed in the supplementary materials of \citep{couvy2020ensemble}.

\subsubsection{Experiments}
Before starting the experiments, we defined a test set by randomly selecting 500 images which respected the same distribution of tiers as the images in the training/validation set. We also verified that the distribution of the manufacturers and the different scanner models was respected. The remaining 5000 images were split into training and validation using a 5-fold cross validation (CV). The separation between training, validation and test sets was made at the patient level to avoid data leakage.
For each of the four tasks considered (SR, gadolinium, tier 3 vs 2-1, tier 2 vs 1), the five models trained in the CV were evaluated on the test set. We also studied the influence of the size of the training set on the performance  by computing learning curves.
We compared the output of each classifier with the consensus label. To set the automatic QC results in perspective, we computed the balanced accuracy (BA) for the raters (defined as the average of the BAs between each rater and the consensus). 

\begin{table}[!t]
\renewcommand{\arraystretch}{1.25}
\begin{center}
    \begin{tabular}{cc}
        \toprule
        \bfseries Characteristics & \bfseries Weighted Cohen's kappa\\
        \hline\hline
        SR (yes vs no) & 0.88\\
        \hline
        Gadolinium injection (yes vs no) & 0.89\\
        \hline
        Contrast (0 vs 1 vs 2) & 0.79\\
        \hline
        Motion (0 vs 1 vs 2) & 0.68\\
        \hline
        Noise (0 vs 1 vs 2) & 0.70\\
        \bottomrule
    \end{tabular}
    \caption{Weighted Cohen's kappa between the two annotators}%
    \label{tab:weighted kappa}
 \end{center}
\end{table}

\section{Results}

\subsection{Manual quality control}

The inter-rater agreement was evaluated using the weighted Cohen's kappa \citep{watson2010method} between the two annotators for each of the characteristics. Results are presented in Table~\ref{tab:weighted kappa}. The agreement is strong for the SR label and the gadolinium injection (0.88 and 0.89) and moderate for the other characteristics (from 0.68 to 0.79).

\begin{figure}[!t]
    \begin{center}
    \includegraphics[width=0.75\linewidth]{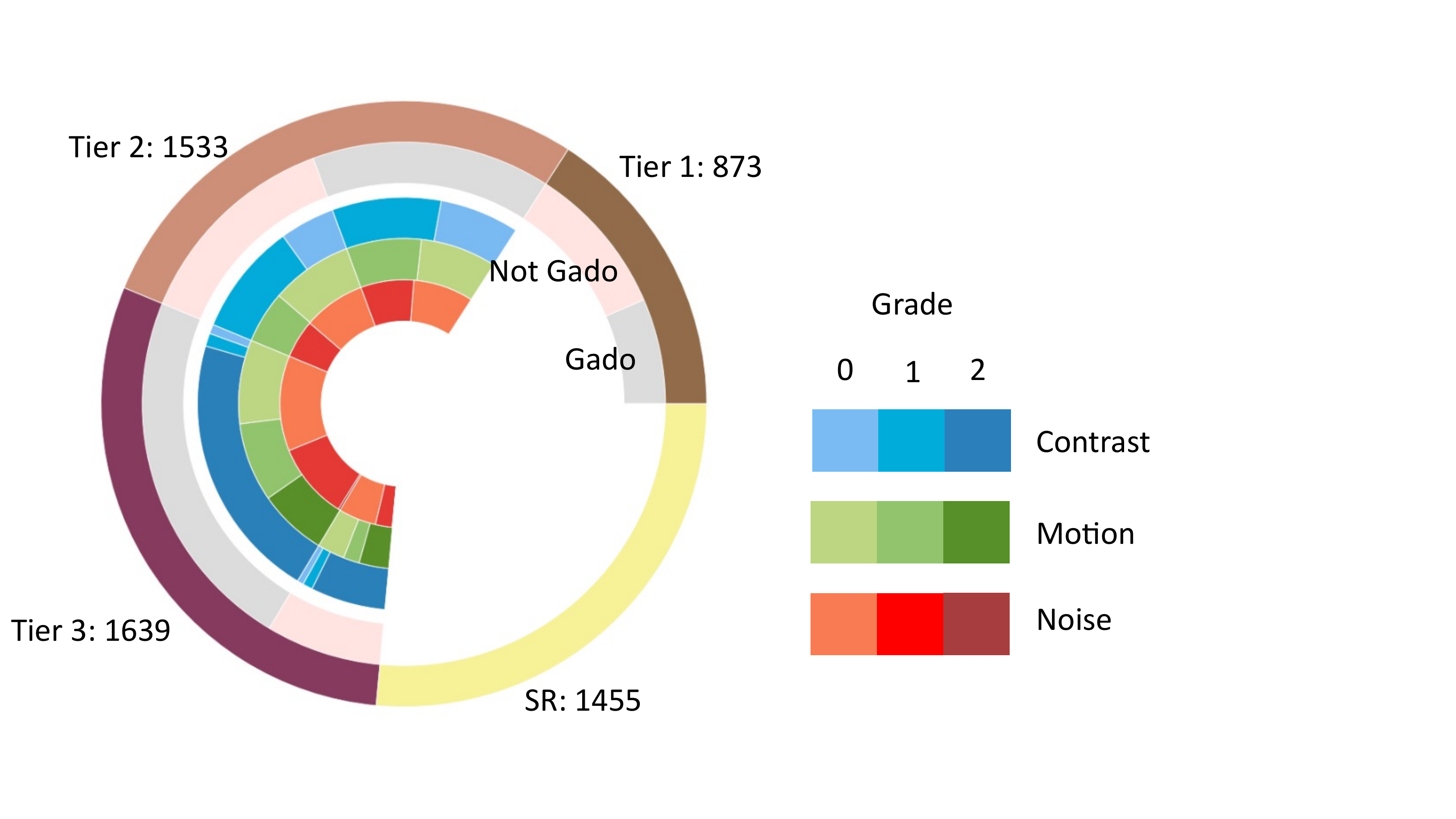}
    \caption{Distribution of the consensus labels for the whole dataset of 5500 images. Outermost circle: images in SR and in the different tiers. For every tier, we divide between images with and without gadolinium injection. For each injection status we see the grade distribution of the contrast, motion and noise characteristics.}
    \label{fig:consensus5500}
    \end{center}
\end{figure}

The distribution of the consensus labels for the 5500 patients is shown in Figure~\ref{fig:consensus5500}. 26\% of the images are labeled as SR, 16\% as tier 1, 28\% as tier 2, and 30\% as tier 3. Figure~\ref{fig:labelbrain} shows some representative examples of T1w brain images with the corresponding labels.

As expected, the proportion of images with gadolinium increased when the quality decreased (proportion of images with gadolinium: 41\% in Tier 1, 53\% in tier 2, 76\% in tier 3; $p<2.13e^{-8}$; $\chi^2$ test). A vast majority of tier 3 images had a contrast of 2 (90\%) and were with gadolinium (70\%). 

If we analyse the relationships between characteristics, we note that 73\% of images with a grade 2 for motion have also a grade 2 for contrast. Unsurprisingly, a strong motion has a severe impact on contrast. On the other hand, images with a grade 2 for contrast present a closer distribution of grade 0, 1 and 2 for motion (40\%, 34\%, and 26\%, respectively).

\begin{figure}[!t]
\begin{center}
    \includegraphics[width=0.8\linewidth]{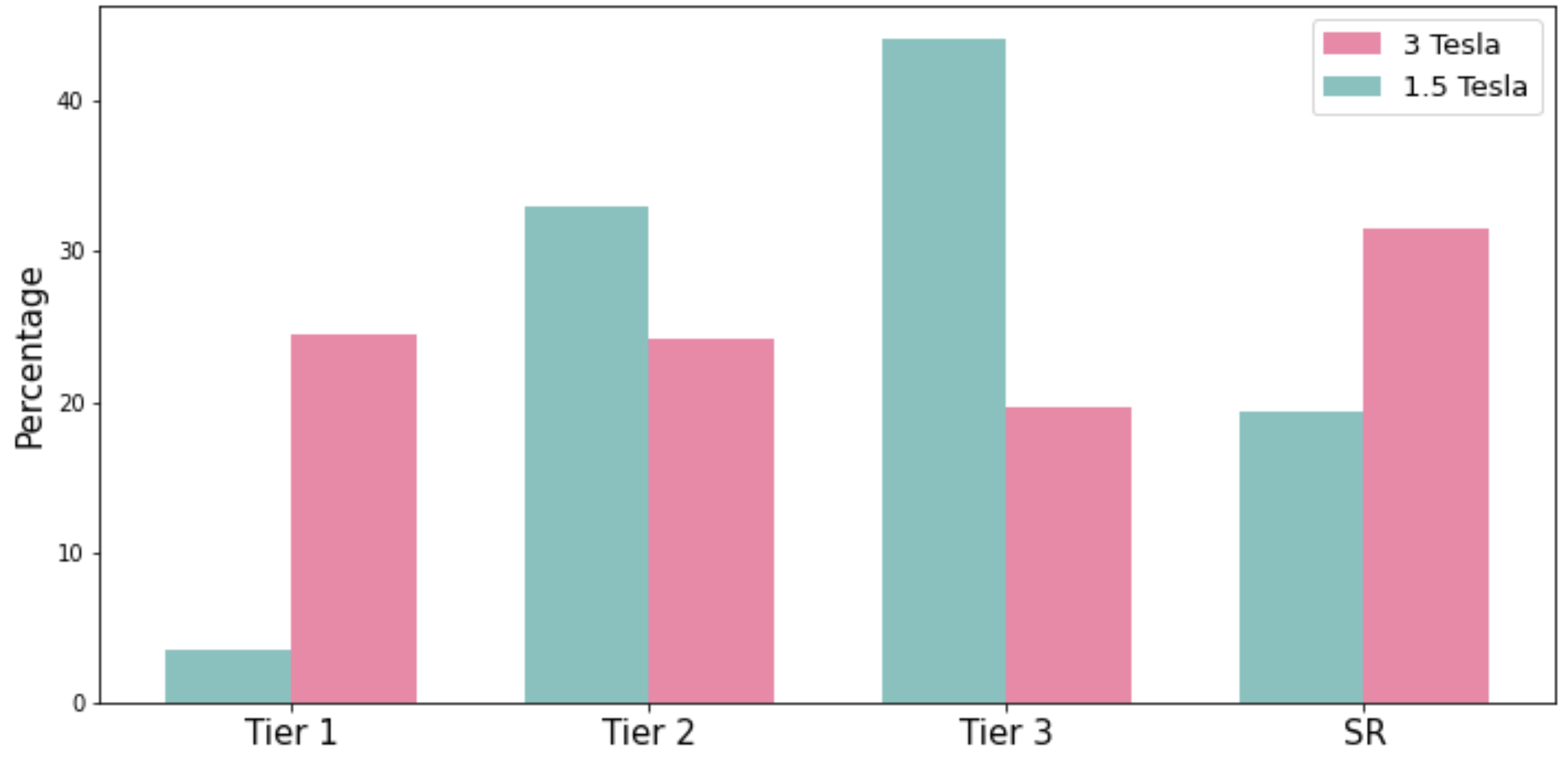}
    \caption{Proportion of images that fall into the different tiers or are labelled as SR depending on the field strength (3T or 1.5T).}
    \label{fig:tesla_consensus}
\end{center}
\end{figure}

Figure~\ref{fig:tesla_consensus} displays the distribution into SR or the different tiers for images acquired at 3T and 1.5T, respectively. One can observe that 3T images are more often in the SR category than 1.5T images (31.5\% vs 19.3\%). The most likely explanation is that 3T scanners are most often equipped with image segmentation tools, which leads to a larger number of segmented images. On the other hand, the image quality tended to be higher for 3T than for 1.5T images, which was expected.

DICOM attributes often contain information regarding the injection of gadolinium. However, it is well-known to radiologists that such information is often unreliable because it is manually entered by the MRI radiographer. We aimed to assess the extent to which such information was unreliable. We thus analysed the ``study description'' and ``series description'' DICOM attributes of the images to check if the presence of gadolinium injection was noted. We considered that it was noted if at least one of the words `gado', `inj' or `iv' was present in the value of one of the attributes. Among the 2416 images that were manually annotated as with gadolinium, 2033 images had the information in the DICOM attributes. Among the 1629 images that were manually annotated as without gadolinium, 987 were noted as images with gadolinium injection according to the DICOM attributes. Since our manual annotation of gadolinium injection is highly reproducible and was designed with the guidance of an experienced neuroradiologist, we conclude that, as expected, DICOM attributes do not provide reliable information regarding the presence of gadolinium. This highlights the importance of being able to detect it using an automatic QC tool.

\subsection{Automatic quality control}
Results obtained for the four tasks of interest by the proposed Conv5\textunderscore FC3 classifier are presented in Table~\ref{tab:results_batch_5500}. We report the BA of the annotators for comparison. For the recognition of SR images, we used all the images available in the training/validation set ($n=5000$); for the gadolinium and tier 3 vs tiers 2-1 tasks, the training/validation set does not include SR images ($n=3770$); and for the tier 2 vs tier 1 task, the training/validation set does not include SR and tier 3 images ($n=2182$).

\begin{table}[!t]
\renewcommand{\arraystretch}{1.25}
    \begin{footnotesize}
        \begin{tabular}{lccccc}
            \toprule
            \bfseries Metric & \makecell{\bfseries SR \\\bfseries(yes vs no)} & \makecell {\bfseries Gadolinium injection \\\bfseries(yes vs no)} & \makecell {\bfseries Tier 3 vs \\\bfseries tiers 2-1} & \makecell {\bfseries Tier 2 vs \\ \bfseries tier 1} \\
            \hline\hline
            BA annotators & 97.13 & 96.10 & 91.56 & 88.27\\\hline
            BA classifiers & 93.76 $\pm$ 0.57 & 97.14 $\pm$ 0.34 & 83.51 $\pm$ 0.93 & 71.65 $\pm$ 2.15 \\\hline
            F1 score & 94.85 $\pm$ 0.41 & 97.04 $\pm$ 0.31 & 84.07 $\pm$ 1.02 & 74.10 $\pm$ 1.35\\\hline
            MCC & 85.71 $\pm$ 1.11 & 94.00 $\pm$ 0.64 & 67.38 $\pm$ 2.13 & 42.10 $\pm$ 3.25\\\hline
            AUC & 93.76 $\pm$ 0.57 & 97.14 $\pm$ 0.34 & 83.51 $\pm$ 0.93 & 71.65 $\pm$ 2.15\\\hline
            Sensitivity & 91.83 $\pm$ 1.18 & 96.45 $\pm$ 0.34 & 79.88 $\pm$ 3.06 & 77.39 $\pm$ 4.29\\\hline
            Specificity & 95.69 $\pm$ 0.53 & 97.82 $\pm$ 0.62 & 87.14 $\pm$ 3.14 & 65.92 $\pm$ 7.47\\\hline
            PPV & 86.44 $\pm$ 1.43 & 98.33 $\pm$ 0.46 & 81.93 $\pm$ 3.36 & 83.20 $\pm$ 2.31\\\hline
            NPV & 97.51 $\pm$ 0.35 &  95.39 $\pm$ 0.42 & 85.83 $\pm$ 1.49 & 57.78 $\pm$ 2.63\\
            \bottomrule
        \end{tabular}
    \end{footnotesize}
    \caption{Results of the CNN classifier for all the tasks. We report the BA of the annotators and for every metric of the CNN we report the mean and the empirical standard deviation across the five folds. BA: balanced accuracy; MCC: Matthews correlation coefficient; AUC: area under the receiver operator characteristic curve; PPV: positive predictive values; NPV: negative predictive values}
    \label{tab:results_batch_5500}
\end{table}

Balanced accuracy for SR and gadolinium is excellent (94\% and 97\%). For SR, the CNN is slightly less good than the annotators. For  gadolinium, the CNN is as good as the raters. For tier 3 vs 2-1, the classifier BA is good but lower than that of the annotators. For tier 2 vs 1, CNN BA is low (71\%) and much lower than that of the raters (88\%).

\begin{figure}[!t]
    \begin{center}
        \includegraphics[width=1\linewidth]{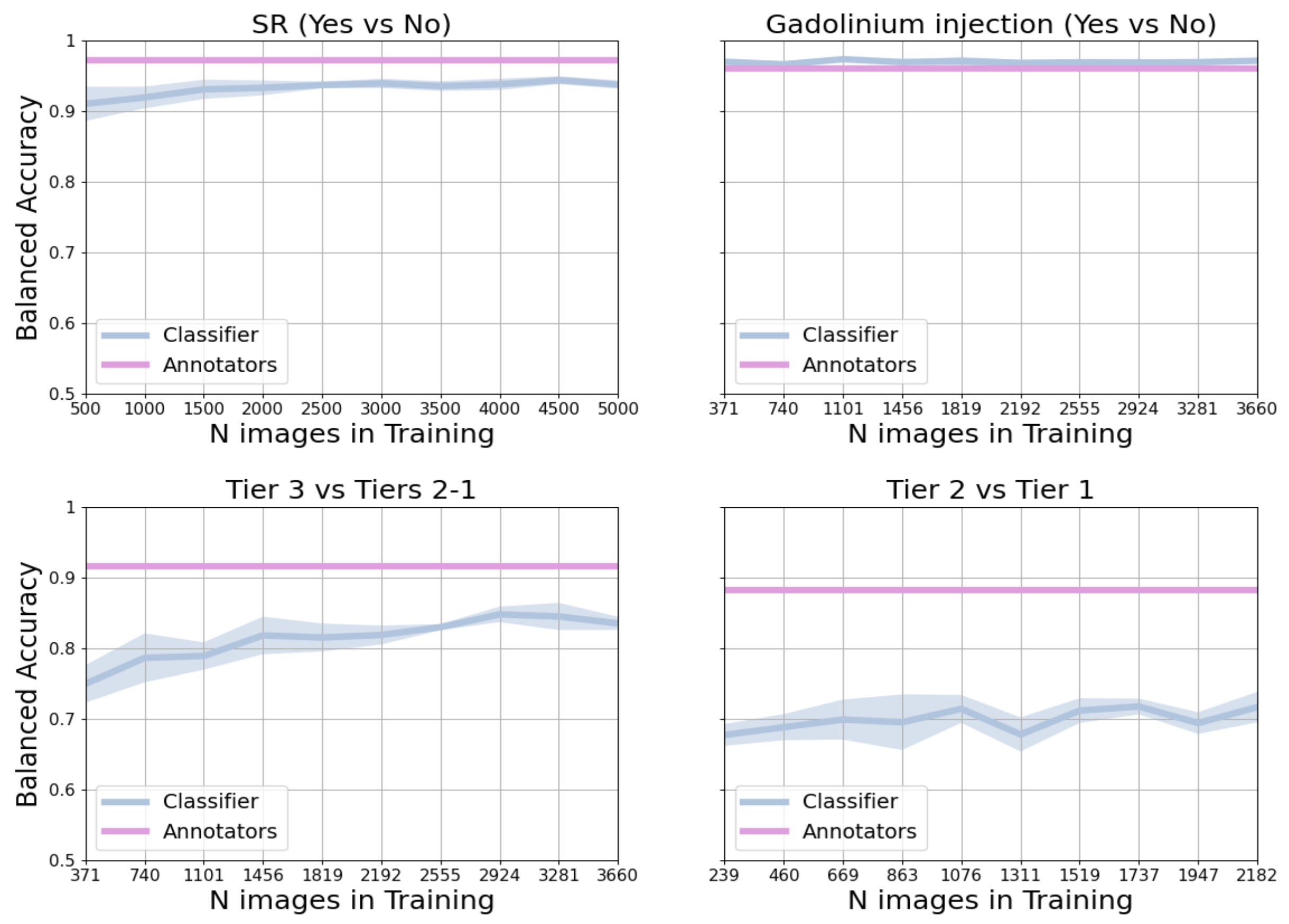}
    \end{center}
    \caption{Learning curves for the SR (yes vs no), gadolinium injection (yes vs no), tier 3 vs tier 2-1 and tier 2 vs tier 1 tasks. Blue: balanced accuracy of the classifier across the five folds. Violet: balanced accuracy of the annotators on the testing set.}
    \label{fig:tier_4_gado_tier3}
\end{figure}

The influence of the size of the training set on the performance is shown in Figure~\ref{fig:tier_4_gado_tier3}. For SR, the performance increases with sample size, even if it is also good with few examples (90\% for 500 images) because of the easiness of the task. For gadolinium, performance is very high regardless of the sample size. For tier 3 vs tiers 2-1, adding more training samples helps the classifier while this is not the case for tier 2 vs 1.

For tier 3 vs tiers 2-1 and tier 2 vs tier 1, we compared the proposed architecture, Conv5\textunderscore FC3, with the Inception and ResNet architectures. For both tasks, the balanced accuracy obtained with the different networks is comparable: while for tier 3 vs tiers 2-1 it is slightly higher with the ResNet (85.82 ± 0.95) than the Conv5\textunderscore FC3 (83.51 ± 0.93) and the Inception (82.40 ± 1.2 ), for tier 2 vs 1 it is slightly higher with the Conv5\textunderscore FC3 (71.65 ± 2.15) than the ResNet (68.08 ± 1.6) or Inception (69.27 ± 2.05) architectures. For both tasks, the performance of the different classifiers were not statistically different (for tier 3 vs tiers 2-1: p\textgreater0.21, McNemar's test; for tier 2 vs tier 1: p\textgreater0.12, McNemar's test).
All the metrics are reported in Table~\ref{tab:inception_resnet}.

\begin{table}[!b]
\begin{center}
\renewcommand{\arraystretch}{1.5}
    \begin{small}
        \bigskip
        \textbf{A. Tier 3 vs tiers 2-1}\\
        \bigskip
        \begin{tabular}{l|ccc}
            \toprule
            \bfseries Metric & \bfseries Conv5\textunderscore FC3 & \bfseries Inception & \bfseries ResNet  \\\hline\hline
            BA          &    83.51 $\pm$ 0.93 & 82.41 $\pm$ 1.28 & 85.82 $\pm$ 0.95 \\\hline
            Sensitivity &    79.88 $\pm$ 3.06 & 75.53 $\pm$ 2.68 & 80.75 $\pm$ 3.24 \\\hline
            Specificity &    87.14 $\pm$ 3.14 & 89.29 $\pm$ 3.45 & 90.89 $\pm$ 2.22 \\\hline
            F1 score &       84.07 $\pm$ 1.02 & 83.38 $\pm$ 1.44 & 86.57 $\pm$ 0.81  \\\hline
            MCC &      67.38 $\pm$ 2.13 & 66.08 $\pm$ 3.02 & 72.52 $\pm$ 1.70  \\\hline
            AUC &      83.51 $\pm$ 0.93 & 82.41 $\pm$ 1.28 & 85.82 $\pm$ 2.81  \\\hline
            PPV &            81.93 $\pm$ 3.36 & 83.80 $\pm$ 3.93 & 86.58 $\pm$ 2.43  \\\hline
            NPV &            85.83 $\pm$ 1.49 & 83.58 $\pm$ 1.20 & 86.85 $\pm$ 1.76  \\
            \bottomrule
        \end{tabular} \\
        \bigskip\bigskip
        \textbf{B. Tier 2 vs tier 1}\\
        \bigskip
        \begin{tabular}{l|ccc}
            \toprule
            \bfseries Metric & \bfseries Conv5\textunderscore FC3 & \bfseries Inception & \bfseries ResNet \\\hline\hline
            BA  &            71.65 $\pm$ 2.15  &  69.28 $\pm$ 2.81  &  68.08 $\pm$ 1.63\\\hline
            Sensitivity &    77.39 $\pm$ 4.29  &  76.86 $\pm$ 4.76  &  82.35 $\pm$ 2.90\\\hline
            Specificity &    65.92 $\pm$ 7.47  &  61.69 $\pm$ 10.01  &  53.80 $\pm$ 4.99\\\hline
            F1 score &       74.10 $\pm$ 1.35  &  72.28 $\pm$ 1.13  &  72.94 $\pm$ 1.18 \\\hline
            MCC &      42.10 $\pm$ 3.25  &  37.74 $\pm$ 4.10  &  37.13 $\pm$ 2.73 \\\hline
            AUC &      71.65 $\pm$ 2.15  &  69.28 $\pm$ 2.81  &  68.08 $\pm$ 1.62 \\\hline
            PPV &            83.20 $\pm$ 2.32  &  81.51 $\pm$ 3.08  &  79.40 $\pm$ 1.34 \\\hline
            NPV &            57.78 $\pm$ 2.63  &  55.49 $\pm$ 1.70  &  58.77 $\pm$ 2.40 \\
            \bottomrule
        \end{tabular}
    \end{small}
    \caption{Results of three 3D CNN architectures (Conv5\textunderscore FC3, Inception and ResNet) for the rating of the overall image quality. We report the mean and the empirical standard deviation across the five folds for all the metrics. BA: balanced accuracy; MCC: Matthews correlation coefficient; AUC: area under the receiver operator characteristic curve; PPV: positive predictive values; NPV: negative predictive values}
    \label{tab:inception_resnet}
\end{center}
\end{table}

\clearpage
\section{Discussion}

In this work, we developed a method for the automatic QC of T1w brain MRI for a large clinical data warehouse. Our approach allows: i) discarding images which are of no interest (SR), ii) recognizing gadolinium injection , iii) rating the overall image quality. 
To this aim, different CNN were trained and evaluated thanks to the manual annotation of 5500 images by two raters.

In the last decades, many computer-aided diagnosis systems using machine learning methods have been proposed for the detection of lesions or tumours, or for the classification of neurodegenerative or psychiatric diseases  \citep{rathore2017review,icsin2016review,burgos2021deep}. Algorithms were mainly developed and tested using research images \citep{samper2018reproducible,noor2019detecting,cuingnet2011automatic}, or clinical datasets of limited size \citep{morin2020accuracy,zhang2019three,campese2019psychiatric,oh2019classification}. Their validation on large realistic clinical datasets is crucial. To that aim, clinical data warehouses, which may gather millions of clinical routine images, offer fantastic opportunities. They also provide considerable challenges. In particular, selecting adequate images for a given analysis task can be very difficult: DICOM attributes may be unreliable, images may be of the wrong type, truncated and their quality is extremely variable. 
Therefore, automatic curation and QC methods are needed to fully exploit the potential of clinical data warehouses. 
Important efforts and achievements have been made by the scientific community to propose protocols and automatic tools for QC. MRIQC \citep{esteban2017mriqc} and VisualQC \citep{raamana2020visual} are two tools developed for the QC of T1w brain MRI data: they propose the extraction of image quality metrics for the detection of outliers, and a graphical interface to check the images. \cite{alfaro2018image} proposed a pipeline for the UK Biobank dataset. \cite{sujit2019automated} trained a CNN using the research dataset ABIDE. Other works focused on QC of processing results (segmentation) rather than raw data \citep{keshavan2018mindcontrol,klapwijk2019qoala}. However, all these tools were designed for research data. Even if the data came from multiple sites, they do not cover all the images existing in a clinical PACS: they did not cover images with gadolinium and the patients presented with a limited number of diseases. On the contrary, in a clinical data warehouse, we may find images with or without gadolinium injection, ``research quality" images, and images segmented, cropped or with so much motion that it is impossible to distinguish the brain. This heterogeneity makes it impossible to use other QC tools present in the literature. To the best of our knowledge, we are the first to propose an automatic QC framework for clinical data warehouses. 

To train our automatic QC algorithm, we had to manually annotate a large sample of images from the data warehouse. It was not possible to use existing protocols and software tools. In addition to the limitations mentioned above, we were also constrained by the environment of the data warehouse which only included a Jupyter notebook and a command-line interface. While constraints may vary from a data warehouse to another, it is very common that the data cannot be downloaded and thus have to be used within a specific informatics set-up \citep{daniel2020hospital}. We thus developed a dedicated visual QC protocol, with the assistance of a resident radiologist. We compared the annotation using 3D images and 2D slices, and we concluded that three 2D slices were sufficient and could represent a good compromise to fulfil our objectives: one being the exclusion of bad quality images that would compromise further analyses.
Manual annotation results showed that our protocol is reproducible across all tasks, even though agreement was weaker for more challenging characteristics. Inter-rater agreement was strong for the SR label and the gadolinium injection and moderate for other characteristics. Manual annotation also provides interesting information on the variability of image quality in a clinical routine data warehouse. As much as 25\% are totally unusable (SR), and almost a third has a very low quality (Tier 3). We also confirmed that gadolinium has a strong impact on image quality, hence the critical importance of detecting it accurately, the DICOM attributes being unreliable in that regard. 

For detecting straight reject, our CNN had excellent performance (BA greater than 90\%). Even though the task is relatively easy, this is very important in order to automatically discard images in a very large scale study. This was also the case for detection of gadolinium, an important characteristic that strongly impacts the behavior of many image analysis methods. For the rating of image quality, the situation was different for identifying Tier 3 (low quality) images and for separating Tier 2 (medium quality) and Tier 1 (high quality). The proposed CNN classifier identified low quality images (Tier 3) with a high accuracy (83\%). This is important because these are typically the images on which image processing algorithms could fail. Differentiating images of high and medium quality could also be useful but is less important as both categories can likely lead to reliable diagnostic predictions. We thus believe that these tools can be reliably used on the rest of this large data warehouse and already have an important practical impact. 
We compared several more sophisticated CNN architectures to our simple network based on five convolutional and three fully connected layers. However, these more complex networks (3D Inception and 3D ResNet) did not provide any significant improvement in performance. 

Thanks to the large number of hospitals in the AP-HP consortium (39 hospitals) and to the huge amount of images collected over the years (1980--now), we strongly believe that this dataset is representative of 3D T1w brain MRI that may be acquired in other hospitals. Consequently, the use of our QC framework could be generalized and it represents a first important step for the use of clinical data warehouses for the design of computer-aided diagnosis systems.

The main limitations of our study concern the annotation process. With the analysis of only three slices, we limit the chances to notice localised artefacts. Another consequence is that it may be difficult to properly distinguish the characteristics when an image is degraded: in particular the motion and the noise may be confused. This is also reflected by moderate values of the weighted Cohen's kappa obtained for these two characteristics.  
Additionally, even if we believe that the CNN models that were trained on data from the AP-HP data warehouse can be applied to other clinical datasets due to the large numbers of hospitals and scanner models involved in study and to the extended period of time, it would be beneficial to apply them on a public dataset for benchmarking.

\section{Conclusion}
In this work, we proposed a framework for the automatic quality control of 3D brain T1w MRI for a large clinical data warehouse. Thanks to the manual annotation of 55O0 images, we trained and validated different convolutional neural networks on 5000 images with a 5-fold CV and we tested them on an independent test set of 500 images. The classifier was as efficient as manual rating for the classification of images which are not proper 3D T1w brain MRI (i.e. truncated or segmented images) and for the images for which gadolinium was injected. In addition, the classifier was able to recognise low quality images with good accuracy.

\clearpage
\singlespacing

\section*{Acknowledgments}
\noindent The research was done using the Clinical Data Warehouse of the Greater Paris University Hospitals. The authors are grateful to the members of the AP-HP WIND and URC teams, and in particular Stéphane Bréant, Florence Tubach, Jacques Ropers, Antoine Rozès, Camille Nevoret, Christel Daniel, Martin Hilka, Yannick Jacob, Julien Dubiel and Cyrina Saussol. They would also like to thank the ``Collégiale de Radiologie of AP-HP'' as well as, more generally, all the radiology departments from AP-HP hospitals. Finally, the authors are very appreciative of the support and guidance they have received from Quentin Vanderbecq when setting up the visual quality control protocol.\\
The research leading to these results has received funding from the Abeona Foundation (project Brain@Scale), from the French government under management of Agence Nationale de la Recherche as part of the ``Investissements d'avenir'' program, reference ANR-19-P3IA-0001 (PRAIRIE 3IA Institute) and reference ANR-10-IAIHU-06 (Agence Nationale de la Recherche-10-IA Institut Hospitalo-Universitaire-6). 

\section*{Authors contribution}
\noindent Study concepts and study design: OC, NB, DD, SB\\
Acquisition, analysis or interpretation of data: all authors\\
Manuscript drafting or manuscript revision for important intellectual content: all authors\\
Approval of final version of submitted manuscript: all authors\\
Literature research: SB, NB, OC\\
Statistical analysis: SB\\
Obtained funding: OC, NB\\
Administrative, technical, or material support:  AM\\
Study supervision: OC, NB, DD\\

\section*{Disclosure statement}
\noindent Competing financial interests related to the present article: none to disclose for all authors.\\
Competing financial interests unrelated to the present article: OC reports having received consulting fees from AskBio (2020), having received fees for writing a lay audience short paper from Expression Santé (2019). Members from his laboratory have co-supervised a PhD thesis with myBrainTechnologies (2016-2019) and with Qynapse (2017-present). OC’s spouse is an employee and holds stock-options of myBrainTechnologies (2015-present). O.C. holds a patent registered at the International Bureau of the World Intellectual Property Organization (PCT/IB2016/0526993, Schiratti J-B, Allassonniere S, Colliot O, Durrleman S, A method for determining the temporal progression of a biological phenomenon and associated methods and devices) (2017).

\section*{APPRIMAGE Study Group}

\noindent Olivier Colliot, Ninon Burgos, Simona Bottani {$^{1}$} \\
Didier Dormont {$^{1,2}$}, Samia Si Smail Belkacem, Sebastian Ströer {$^{2}$}\\
Nathalie Boddaert {$^{3}$} \\
Farida Benoudiba, Ghaida Nasser, Claire Ancelet, Laurent Spelle {$^{4}$}\\
Hubert Ducou-Le-Pointe{$^{5}$}\\
Catherine Adamsbaum{$^{6}$}\\
Marianne Alison{$^{7}$}\\
Emmanuel Houdart{$^{8}$}\\
Robert Carlier {$^{9,17}$}\\
Myriam Edjlali{$^{9}$}\\
Betty Marro{$^{10,11}$}\\
Lionel Arrive{$^{10}$}\\
Alain Luciani{$^{12}$}\\
Antoine Khalil{$^{13}$}\\ 
Elisabeth Dion{$^{14}$}\\
Laurence Rocher{$^{15}$}\\
Pierre-Yves Brillet{$^{16}$}\\ 
Paul Legmann, Jean-Luc Drape {$^{18}$}\\ 
Aurélien Maire, Stéphane Bréant, Christel Daniel, Martin Hilka, Yannick Jacob, Julien Dubiel, Cyrina Saussol {$^{19}$}\\
Florence Tubach, Jacques Ropers, Antoine Rozès, Camille Nevoret {$^{20}$}\\

\begin{small}
\noindent $^{1}$ Paris Brain Institute (ICM), Inserm U 1127, CNRS UMR 7225, Sorbonne Université, Inria, Aramis project-team, F-75013, Paris, France \\
$^{2}$  AP-HP, Hôpital de la Pitié Salpêtrière, Department of Neuroradiology, F-75013, Paris, France \\
$^{3}$  AP-HP, Hôpital Necker, Department of Radiology, F-75015, Paris, France \\
$^{4}$  AP-HP, Hôpital Bicêtre, Department of Radiology, F-94270, Le Kremlin-Bicêtre, France \\
$^{5}$  AP-HP, Hôpital Armand-Trousseau, Department of Radiology, F-75012, Paris, France \\
$^{6}$  AP-HP, Hôpital Bicêtre, Department of Pediatric Radiology, F-94270, Le Kremlin-Bicêtre, France \\
$^{7}$  AP-HP, Hôpital Robert-Debré, Department of Radiology, F-75019, Paris, France \\ 
$^{8}$  AP-HP, Hôpital Lariboisière , Department of Neuroradiology, F-75010, Paris, France \\
$^{9}$  AP-HP, Hôpital Raymond-Poincaré, Department of Radiology, F-92380, Garches, France \\
$^{10}$ AP-HP, Hôpital Saint-Antoine, Department of Radiology, F-75012, Paris, France \\
$^{11}$ AP-HP, Hôpital Tenon, Department of Radiology, F-75020, Paris, France \\
$^{12}$ AP-HP, Hôpital Henri-Mondor, Department of Radiology, F-94000, Créteil, France \\
$^{13}$ AP-HP, Hôpital Bichat, Department of Radiology, F-75018, Paris, France \\
$^{14}$ AP-HP, Hôpital Hôtel-Dieu, Department of Radiology, F-75004, Paris, France \\
$^{15}$ AP-HP, Hôpital Antoine-Béclère, Department of Radiology, F-92140, Clamart, France \\
$^{16}$ AP-HP, Hôpital Avicenne, Department of Radiology, F-93000, Bobigny, France \\
$^{17}$ AP-HP, Hôpital Ambroise Paré, Department of Radiology, F-92100 104, Boulogne-Billancourt, France \\ 
$^{18}$ AP-HP, Hôpital Cochin, Department of Radiology, F-75014, Paris, France \\ 
$^{19}$ AP-HP, WIND department, F-75012, Paris, France \\
$^{20}$ AP-HP, Unité de Recherche Clinique, Hôpital de la Pitié Salpêtrière, Department of Neuroradiology, F-75013, Paris, France  \\
\end{small}

\newpage
\bibliographystyle{elsarticle-harv} 
\bibliography{QC}

\clearpage
\setcounter{page}{1}
\beginsupplement
\begin{center}
    \begin{Large}
    Automatic quality control of brain T1-weighted magnetic resonance images for a clinical data warehouse \bigskip
    \end{Large}
    
    Simona Bottani$^{a,b,c,d,e,f}$, Ninon Burgos$^{b,c,d,e,f,a}$, Aurélien Maire$^g$, Adam Wild$^{b,c,d,e,f,a}$, Sebastian Ströer$^h$, Didier Dormont$^{b,c,d,e,f,a,h}$, Olivier Colliot$^{b,c,d,e,f,a}$, APPRIMAGE Study Group \bigskip
    
    $^a$Inria, Aramis project-team, Paris, 75013, France\\
    $^b$Sorbonne Université, Paris, 75013, France \\
    $^c$Institut du Cerveau - Paris Brain Institute - ICM, Paris, 75013, France\\
    $^d$Inserm, Paris, 75013, France\\
    $^e$CNRS, Paris, 75013, France\\
    $^f$AP-HP, Hôpital de la Pitié Salpêtrière, Paris, 75013, France\\
    $^g$AP-HP, WIND department, Paris, 75012, France \\
    $^h$AP-HP, Hôpital de la Pitié Salpêtrière, Department of Neuroradiology, Paris, 75013, France\\
    
    \bigskip
    
    - \bigskip
    
    \begin{Large}
    Supplementary Material \bigskip
    \end{Large}
\end{center}
\clearpage

\begin{table}[h]
\begin{center}
\renewcommand{\arraystretch}{1.1}
\begin{small}
    \begin{tabular}{cccc}
        \toprule
        \bfseries Manufacturer & \bfseries Model Name & \bfseries Field strength [T]  & \bfseries N images\\\hline\hline
        \multirow{12}{*}{Siemens  Healthineers}     
            & Aera                          & 1.5 & 489\\\cline{2-4}
            & Amira                         & 1.5 & 29\\\cline{2-4}
            & Avanto                        & 1.5 & 603\\\cline{2-4}
            & Avanto\textunderscore fit     & 1.5 & 81\\\cline{2-4}
            & Biograph mMR                  & 3 & 12\\\cline{2-4}
            & Espree                        & 1.5 & 1\\\cline{2-4}
            & Magnetom Vida                 & 3 & 3\\\cline{2-4}
            & Magnetom Essenza              & 1.5 & 11\\\cline{2-4}
            & Sempra                        & 1.5 & 3\\\cline{2-4}
            & Skyra                         & 3 & 1851\\\cline{2-4}
            & Spectra                       & 3 & 23\\\cline{2-4}
            & Sympthony                     & 1.5 & 3\\\cline{2-4}
            & Verio                         & 3 & 643\\\hline
        \multirow{12}{*}{GE  Healthcare}            
            & Discovery MR450               & 1.5 & 4\\\cline{2-4}
            & Discovery MR750(w)            & 3 & 675\\\cline{2-4}
            & Optima MR360                  & 1.5 & 2\\\cline{2-4}
            & Optima MR450w                 & 1.5 & 284\\\cline{2-4}
            & Signa Architect               & 1.5 & 243\\\cline{2-4}
            & Signa Artist                  & 1.5 & 4\\\cline{2-4}
            & Signa Excite                  & 1.5 & 3\\\cline{2-4}
            & Signa Explorer                & 1.5 & 1\\\cline{2-4}
            & Signa HDx(t)                  & 1.5 & 489\\\cline{2-4}
            & Signa Pioneer                 & 3 & 1\\\cline{2-4}
            & Signa Voyager                 & 1.5 & 1\\\cline{2-4}
            & Unknown                       & 1.5 & 3\\\hline
        \multirow{3}{*}{Philips}                 
            & Achieva                       & 3 & 21\\\cline{2-4}
            & Ingenia                       & 1.5 & 5\\\cline{2-4}
            & Intera                        & 1.5 & 7\\\hline
        \multirow{2}{*}{Toshiba}                  
            & Titan                         & 1.5 & 2\\\cline{2-4}
            & Vantage Elan                  & 1.5 & 3\\
        \bottomrule
    \end{tabular}
    \caption{Model name of all the scanners, grouped by the manufacturer, with the corresponding magnetic field strength [T] and the number of images.}
    \label{tab:machines_name}
\end{small}
\end{center}
\end{table}
\clearpage

\begin{sidewaystable}
\begin{center}
\begin{small}
\renewcommand{\arraystretch}{1.25}
    \begin{tabular}{ccccccc}
        \toprule
        \bfseries Layer & \bfseries Filter size & \makecell{\bfseries Number of \\\bfseries filters/\\\bfseries neurons} &  \makecell{\bfseries Stride \\\bfseries size} & \makecell{\bfseries Padding \\\bfseries size} & \makecell{\bfseries Dropout \\\bfseries rate} & \makecell{\bfseries Output\\\bfseries size} \\
        \hline\hline
        
        Conv+BN+ReLU - 1    & 3x3x3     & 8     & 1     & 1           & --    & 8x169x208x179\\\hline
        MaxPool - 1         & 2x2x2     & --    & 2     & adaptive    & --    & 8x85x104x90\\\hline
        Conv+BN+ReLU - 2    & 3x3x3     & 16    & 1     & 1           & --    & 16x85x104x90\\\hline
        MaxPool - 2         & 2x2x2     & --    & 2     & adaptive    & --    & 16x43x52x45\\\hline
        Conv+BN+ReLU - 3    & 3x3x3     & 32    & 1     & 1           & --    & 32x43x52x45\\\hline
        MaxPool - 3         & 2x2x2     & --    & 2     & adaptive    & --    & 32x22x26x23\\\hline
        Conv+BN+ReLU - 4    & 3x3x3     & 64    & 1     & 1           & --    & 64x22x26x23\\\hline
        MaxPool - 4         & 2x2x2     & --    & 2     & adaptive    & --    & 64x11x13x1\\\hline
        Conv+BN+ReLU - 5    & 3x3x3     & 128   & 1     & 1           & --    & 128x11x13x12\\\hline
        MaxPool - 5         & 2x2x2     & --    & 2     & adaptive    & --    & 128x6x7x6\\\hline
        Dropuout            & --        & --    & --    & --          & 0.5   & 128x6x7x6\\\hline
        FC - 1              & --        & 1300  & --    & --          & --    & 1500\\\hline
        FC - 2              & --        & 50    & --    & --          & --    & 50\\\hline
        FC - 3              & --        & 2     & --    & --          & --    & 2\\\hline
        Softmax             & --        & --    & --    & --          & --    & 2\\
        
        \bottomrule
    \end{tabular}
    \caption{Hyperparameters of the 3D Conv5\textunderscore FC3 CNN. BN: batch normalization; Conv: convolutional layer; FC: fully connected; MaxPool: max pooling.}
    \label{tab:parameters}
\end{small}
\end{center}
\end{sidewaystable}
\clearpage

\end{document}